\documentclass[12pt]{article}
\usepackage[utf8]{inputenc}
\usepackage{setspace}
\AtBeginEnvironment{table}{\singlespacing}
\usepackage[margin=1.0in]{geometry}
\usepackage{graphicx}
\graphicspath{ {./figures/} }
\usepackage{amsmath}
\usepackage{amsfonts}
\usepackage{dsfont}
\usepackage{lineno}
\usepackage{xcolor}
\usepackage{hyperref}
\usepackage[document]{ragged2e}
\usepackage{subcaption}
\usepackage{algpseudocode}
\usepackage{algorithm}
\usepackage{nameref}
\usepackage[T1]{fontenc}
\usepackage[french,english]{babel}
\DeclareSymbolFont{bbold}{U}{bbold}{m}{n}
\DeclareSymbolFontAlphabet{\mathbbold}{bbold}

\usepackage{etoolbox}
\AtBeginEnvironment{figure}{\setlength{\RaggedRightParindent}{0em}}
\AtBeginEnvironment{table}{\setlength{\RaggedRightParindent}{0em}}

\usepackage{csquotes}
\usepackage[style=authoryear-ibid,backend=biber]{biblatex}
\usepackage{array}
\usepackage{makecell}

\newcommand{\ind}{\perp\!\!\!\!\perp}

\usepackage{titling}
\predate{}
\postdate{}
\usepackage{authblk}
\title{Small area estimation of forest biomass via a two-stage model for continuous zero-inflated data}
\date{} 
\author[1,2,*]{Grayson W. White} 
\author[3]{Josh K. Yamamoto} 
\author[4]{Dinan H. Elsyad} 
\author[5]{Julian F. Schmitt} 
\author[4]{Niels H. Korsgaard} 
\author[6]{Jie Kate Hu} 
\author[7]{George C. Gaines, III} 
\author[8]{Tracey S. Frescino} 
\author[4]{Kelly S. McConville} 
\affil[1]{{\small Department of Forestry, Michigan State University, East Lansing, MI, USA}}
\affil[2]{{\small Department of Statistics \& Probability, Michigan State University, East Lansing, MI, USA}}
\affil[3]{{\small Redcastle Resources, Inc., Salt Lake City, UT, USA}}
\affil[4]{{\small Department of Statistics, Harvard University, Cambridge, MA, USA}}
\affil[5]{{\small Environmental Science and Engineering, California Institute of Technology, Pasadena, CA, USA}}
\affil[6]{{\small Department of Statistics, The Ohio State University, Columbus, OH, USA}}
\affil[7]{{\small Rocky Mountain Research Station Forest Inventory \& Analysis, USDA Forest Service, Missoula, MT, USA}}
\affil[8]{{\small Rocky Mountain Research Station Forest Inventory \& Analysis, USDA Forest Service, Riverdale, UT, USA}}
\affil[*]{{\small Corresponding author: Grayson W. White, whitegra@msu.edu}}

\begin{document}
\maketitle

\newpage
\begin{abstract} 
The United States (US) Forest Inventory \& Analysis Program (FIA) collects data on and monitors the trends of forests in the US. FIA is increasingly interested in monitoring forest attributes such as biomass at fine geographic and temporal scales, resulting in a need for assessment and development of small area estimation techniques in forest inventory. We implement a small area estimator and parametric bootstrap estimator that account for zero-inflation in biomass data via a two-stage model-based approach and compare its performance to a post-stratified estimator and to the unit- and area-level empirical best linear unbiased prediction (EBLUP) estimators. For estimator comparison, we conduct a simulation study with counties in the US state Nevada as domains based on sampled plot data and remote sensing data products. Results show the zero-inflated estimator has the lowest relative bias and the smallest empirical root mean square error. Moreover, the 95\% confidence interval coverages of the zero-inflated estimator and the unit-level EBLUP are more accurate than the other two estimators. To further illustrate the practical utility, we employ a data application across the 2019 measurement year in Nevada. We introduce the \texttt{R} package, \texttt{saeczi}, which efficiently implements the zero-inflated estimator and its mean squared error estimator. 
\end{abstract}

\selectlanguage{french} 
\begin{abstract}
Le programme d’analyse et d’inventaire forestier (AIF) des États-Unis recueille les données forestières et surveille de près l’état des forêts aux États-Unis.  L’intérêt croissant de l’AIF dans le suivi des caractéristiques des forêts, telles que la biomasse sur des échelles géographiques et temporelles fines, engendre la nécessité de développer des techniques d’estimation pour de petites régions géographiques adaptées au contexte des inventaires forestiers.  Nous mettons en œuvre un estimateur pour petites régions géographiques ainsi qu’un estimateur paramétrique par rééchantillonnage (<< Bootstrap >>) qui prennent en compte l’excès de zéros dans les données de biomasse grâce à une approche en deux étapes fondée sur des modèles. Nous comparons la performance de cet estimateur à l’estimateur post-stratifié et au meilleur prédicteur linéaire sans biais empirique (MPLSBE) aux niveaux unitaires et régionaux. Nous réalisons une étude de simulation dans les comtés du Nevada, aux États-Unis, en se basant sur des données réelles obtenus par échantillonnage de placettes et de produits de données de télédétection. Les résultats montrent que l’estimateur avec excès de zéros présente le biais relatif le plus faible et la plus petite erreur quadratique moyenne empirique. De plus, les couvertures des intervalles de confiance à 95\% de l’estimateur avec excès de zéros et du MPLSBE au niveau unitaire sont plus précis que celles des deux autres estimateurs. Afin d’illustrer davantage l’aspect pratique, nous réalisons une application de données à partir des observations du Nevada de l’année 2019. Nous introduisons également le package en R, \texttt{saeczi}, qui met en œuvre l’estimateur avec excès de zéros et son estimateur quadratique de manière efficiente.
\end{abstract}
\selectlanguage{english} 

{ 
\small \textbf{Keywords:} small area estimation, forest inventory and analysis, zero-inflated models, biomass estimation, remote sensing, statistical software
}

\newpage

\section{Introduction}\label{sec:intro}

The United States Forest Service Forest Inventory \& Analysis (FIA) Program is the strategic nationwide forest inventory (NFI) of the United States (US). FIA collects and analyzes data to evaluate and report on the extent and status of, as well as time-varying trends in, the forests of the US. These data consist primarily of measurements of forest vegetation traits made at permanent sample locations distributed across most of the nation and selected according to an equal probability, quasi-systematic sample design. The design prescribes superimposition of a grid of approximately 2,400 hectare hexagons across the US, and selects one plot location at random within each hexagon to sample. Another important feature of the design is the system of time-interpenetrating panels (10 in the western US, and 5 or 7 in the eastern US) which govern inventory cycles.

The FIA sample design permits unbiased direct estimation of forest population parameters, and estimation of associated variances, in a design-based inferential framework. The FIA design produces adequately precise design-based estimates for populations of historic importance to FIA such as, for example, US state-level estimates over an inventory cycle-wide time period. However, design-based estimates for subpopulations spanning sub-state geographies and/or short time intervals are often too unstable to permit meaningful inferences on forest attributes of interest.


These imprecise design-based estimates for subpopulations are particularly relevant due to increased interest in forest attribute estimates in sub-state regions such as counties, ecological regions and subdivisions thereof, watersheds, recently burned regions, and more. This increased interest motivated extensive research into NFI applications of small area estimation (SAE) methods (see \textcite{rao15} for SAE methods) \parencite{wiener2021united}. The resounding demand for the FIA-based SAE methods and tools was acknowledged by law in recent appropriations bills, wherein US Congress directed FIA ``...to further explore the use of available technologies like remote sensing and methodologies such as small area estimation to further refine county and State biomass estimates as outlined in Sec. 8632 of the Agriculture Improvement Act of 2018'' \parencite{congress2022}.

FIA's standard estimator is a direct, design-based and model-assisted, post-stratified estimator \parencite{green_book}. This estimator combines plot-level data with a single categorical auxiliary variable and can be represented as a weighted sum of sample means within post-strata. 

When these post-strata are homogeneous within- and heterogeneous between- post-strata, the post-stratified estimator allows for variance reduction over a Horvitz-Thompson estimator. However, as mentioned above, this estimator provides insufficient precision for small areas of interest with few numbers of plots. Further explorations of design-based, model-assisted estimators, such as the generalized regression estimator, still exhibit insufficient precision in small sample size scenarios in forest inventory \parencite{Frescino2022}. 

The instability of the direct estimator described above highlights the need to consider alternative approaches to estimation in these small areas, and in particular, the viability of a \textit{model-based} and \textit{indirect} estimators. For reference, \textit{indirect} estimators consider data outside of the domain of interest to help inform estimates, while \textit{direct} estimators, such as the post-stratified estimator, only consider data in the domain of interest; further the \textit{model-based} inferential paradigm relies on distributional assumptions of the proposed model, while the \textit{design-based} inferential paradigm relies only on the design of the sample. Two common indirect, model-based approaches in small sample size cases are the empirical best linear unbiased prediction (EBLUP) estimators based on the Fay-Herriot and Battese-Harter-Fuller models \parencite{fay1979estimates, battese1988error}, which are fit at the area- and unit-level, respectively, but are analogous in the sense that each of them implements an area-specific random intercept which captures the between-area variation of the data. In forest inventory estimation applications, it is common to use auxiliary information derived from aerial and satellite-borne remote sensing platforms, including radiometric, climatic, topographic, canopy height, and other data products to inform the model. Investigations of these model forms feature prominently in the remote sensing-based forest inventory literature \parencite{breidenbach2012small, ver2018hierarchical, temesgen2021using, cao2022increased, Frescino2022}, however despite precision gains shown for small area estimates, they can exhibit significant limitations in forest inventory applications. 

Oftentimes, even with the rich auxiliary data available to use in the modern day, the EBLUP estimator based on the Battese-Harter-Fuller model is poorly specified as the relationship between the forest inventory data and this auxiliary data causes modeling assumptions to not be met (see, e.g., \textcite{Frescino2022} Figure 9), in turn causing inaccurate estimates. This model misspecification is often caused in part by zero-inflation in forest attribute response variables. 

As the name suggests, data are canonically classified as being zero-inflated when they contain a significant proportion of zeroes. While it's hardly ever very productive to spell out a definition for a phrase that is its own definition, we do so here to emphasize the fact that to call data zero-inflated is to only say something very broad about how that data is distributed. There is no commonly accepted cutoff for at what proportion of zeros our data deserves the label zero-inflation, and there is no restriction on the distribution of the non-zero data. While the work done in this paper concerns zero-inflated data with no constraint on the level of ``zeroness'', we do require that the non-zero data is positive and continuously distributed. Moreover, as we are working in an estimation setting, when we say our data is zero-inflated we mean that the \textit{response variable} is zero-inflated. In the forest inventory context, FIA collected forest attribute variables often exhibit this specific type of zero-inflation due to the moderate frequency at which there is a lack of trees and vegetation at plot locations.

The difficulty of specifying a model which meets modeling assumptions at the unit-level can sometimes lead researchers to take an area-level approach such as the EBLUP estimator based on the Fay-Herriot model \parencite{white2021, may2023}. In forest inventory applications, the relationship between forest attributes of interest and auxiliary data is often stronger and more linear at the area-level. However, area-level models substantially reduce the number of data points and do not take full advantage of the auxiliary data available at fine resolutions. Further, the EBLUP estimator based on the Fay-Herriot model treats the within-area variation as a fixed and known quantity, while in reality this variation is an estimate derived from the unit-level sample. This assumption is particularly troublesome in small area applications, where the data used to inform the within-area variation is limited.

The limitations of these estimators motivate an approach that achieves the ``best of both worlds'': a model-based approach that takes advantage of rich unit-level data and meets modeling assumptions. A two-stage modeling approach for SAE that accounts for zero-inflation at the unit-level was first proposed and implemented by \textcite{pfeffermann}. This method first specifies a linear mixed model fit to the positive response values in the sample data, and then a generalized linear mixed model with binomial response fit to the entire sample data to classify the probability of non-zero for a given unit. These models were fit in a Bayesian framework with Markov chain Monte Carlo (MCMC) simulations. \textcite{chandra_sud} propose a similar modeling approach in a frequentist framework, and estimate the mean squared error of small area predictions through a parametric bootstrap approach. Similar zero-inflated estimators have been applied to forest inventory data as well. In particular, \textcite{finley2011hierarchical} propose a Bayesian two-stage model-based approach with spatial random effects to account for zero-inflation in forest inventory data for pixel-level prediction of forest attributes of interest, however they do not implement or assess small area estimates. 

Here, we implement the two-stage frequentist small area estimator (henceforth zero-inflated estimator) and parametric bootstrap algorithm as described in \textcite{chandra_sud} to predict average live above-ground square root biomass (henceforth average square root biomass) in small areas of interest. In Section \ref{sec:methods} we introduce the response and auxiliary data, define the estimators used for this analysis, and discuss our simulation framework. In Section \ref{sec:sim-res}, we conduct a simulation study in the US state of Nevada (henceforth Nevada), using FIA plot data and a variety of remotely-sensed data products with Nevada's counties as our small areas of interest. The counties have a large proportion of plots with no biomass, ranging from 58\% to 98\% zero. Next, in Section \ref{sec:data-application}, we apply these estimators to predict average square root biomass in one measurement year in Nevada, where each county defines a small area of interest. In both Sections \ref{sec:sim-res} and \ref{sec:data-application}, we compare the zero-inflated estimator to the EBLUP estimators based on the Battese-Harter-Fuller and Fay-Herriot models, and to FIA's standard post-stratified estimator. Finally, in Section \ref{sec:discussion} we describe the technical implementation of the zero-inflated estimator through the introduction of our \texttt{R} package, \texttt{saeczi}, and discuss the broader impacts and implications of our work. 

\section{Methods}\label{sec:methods}

\subsection{Data}\label{sec:data}

Throughout this article, our goal is to estimate average square root biomass in a variety of small areas of interest, and to assess the quality of the estimators used. As a case study, we focus on average square root biomass estimation in the counties of Nevada as our response variable. The data for our response variable come from ground plots collected by the USDA Forest Service FIA Program, with sampling design as described in Section \ref{sec:intro}. Note that the choice to model and predict average \textit{square root} biomass is discussed in Appendix \ref{appendix:diagnostics}. Further, we have wall-to-wall auxiliary data from a variety of remotely-sensed and climatic raster files, which have all been resampled to a $90\text{m} \times 90\text{m}$ pixel size. For our analyses, we focus on 19 auxiliary variables as listed and described in Table~\ref{tab:XVariables}. The FIA ground plots and auxiliary data were extracted and matched with the \texttt{FIESTA} \texttt{R} package \parencite{frescino2023fiesta}. 

The dataset used in Section~\ref{sec:sim-res} contains 11,848 FIA plot observations across the 17 counties in Nevada, linked with 19 auxiliary variables listed in Table~\ref{tab:XVariables}. We used the most current sampled measurement for each plot in the FIA database downloaded on February 8, 2023 \parencite{fiadata}. For the analyses in Section~\ref{sec:sim-res} we do not need pixel-level data in Nevada, because in the simulation study we treat the entire set of plots as our target population and draw Monte Carlo samples from this set. This approach is described in detail in Section~\ref{sec:sim-study}.

For the data application in Section~\ref{sec:data-application} we used a subset of the data described above to demonstrate estimator efficacy at constrained temporal scales. In particular, we used only the FIA plot observations from measurement year 2019, resulting in a sample dataset of just 1,155 observations. Considering only FIA measurements made in 2019, Nevada county sample sizes range from 3 to 190 observations with a median of 60 observations. Further, for our analysis we have access to each of the auxiliary variables described in Table~\ref{tab:XVariables} across the 33.7 million 90m pixels comprising Nevada, our population of interest.  

\subsection{Estimators}\label{sec:estimators}

\subsubsection{Notation}

Before describing the estimators, we introduce our notation. Let the indexed set $U = \{1, 2, \dots, N \}$ denote a finite population with $N$ units. $U$ can be partitioned into $J$ small areas indexed by $j$ where each small area population, $U_j$, contains $N_j$ units. Let $s \subset U$ denote our sample of $n$ units and the subset of sampled units $s_j$ represent the sampled units in small area $j$. Finally, the individual units are indexed by $i$. 

We use $y$ to denote the response variable of interest, square root biomass, where $y_{ij}$ represents the observed value for the $i^{\text{th}}$ unit in small area $j$. We are interested in estimating $\mu_j = \frac{1}{N_j}\sum_{i\in U_j}y_{ij},$ the population mean of square root biomass in small area $j$.  Since we are interested in zero-inflation in $y$, we also introduce $z_{ij}$, a binary variable indicating whether $y_{ij}$ is positive, formally
$$
    z_{ij} = \begin{cases} 
      1 & \text{if} \ \ y_{ij} > 0, \\
      0 & \text{if} \ \ y_{ij} = 0.
   \end{cases}
$$

Further, for modeling we denote the unit-level vector of predictors by 
$$
    \mathbf{x}_{ij} = (1,~ x_{1_{ij}},~ x_{2_{ij}}, \dots,~ x_{p-1_{ij}})^T,
$$
containing the intercept and $p-1$ predictor values for the $j$th county's $i$th plot. At the area-level, we denote the vector of predictors by 
$$
    \mathbf{\overline X}_{j} = (1,~ \overline X_{1_{j}},~ \overline X_{2_{j}}, \dots,~ \overline X_{p-1_{j}})^T,
$$
where each predictor in $\mathbf{\overline X}_{\#_j}$ is the population mean for that predictor in a given small area. We use this quantity for both model fitting and prediction. For prediction with the zero-inflated estimator and unit-level EBLUP estimator, we use $\mathbf{X}_{ij}$ to denote the vector analogous to $\mathbf{x}_{ij}$ with the subtle difference that this is a vector of population units.

\subsubsection{Horvitz-Thompson}
Due to its reference throughout our analyses, we first introduce the Horvitz-Thompson estimator of $\mu_j$ \parencite{horvitz1952generalization}, which is given by

\begin{equation}
    \hat{\mu}_j^{\text{HT}} = \frac{1}{n_j}\sum_{i\in s_j}y_{i}.
\end{equation}

The Horvitz-Thompson estimator does not take into account any auxiliary data and simplifies to the sample mean here because of our equal-probability sampling design. The mean square error estimator is given by

$$
    \widehat{\text{MSE}}\Big(\hat{\mu}_j^{\text{HT}}\Big) = \frac{1}{n_j(n_j-1)}\sum_{i\in s_j}\Big(y_i - \hat{\mu}_j^{\text{HT}}\Big)^2,
$$
as described in \textcite{sarndal2003model} when we ignore the finite population correction term.  Because the Horvitz-Thompson estimator is approximately unbiased under equal probability designs, its variance and mean squared error are roughly equivalent \parencite{sarndal2003model}.  Therefore, we present the uncertainty error in terms of a MSE estimator instead of a variance estimator to be consistent with the model-based estimators presented in this article.

\subsubsection{Post-Stratified}

As mentioned in Section~\ref{sec:intro}, FIA still often relies on the direct, model-assisted, post-stratified estimator for production processing of its estimates. The post-stratified estimator is the result of taking a weighted sum of Horvitz-Thompson estimates that are built over the $H$ post-strata of a single categorical auxiliary variable within a small area, with individual post-strata indexed by $h$. Each weight is simply the proportion of population units in that small area that belong to the corresponding post-stratum. In our case, we use the binary categorical auxiliary variable that classifies each unit in the population into either a forested or non-forested post-stratum, so $H = 2$. The post-stratified estimator \parencite{holt1979post} is as follows
$$
    \hat{\mu}_{j}^{\text{PS}} = \frac{1}{N_j}\sum_{h = 1}^2 \frac{N_{jh}}{n_{jh}}\sum_{i\in s_{jh}}y_i = \sum_{h = 1}^2\frac{N_{jh}}{N_j}\hat{\mu}_{jh}^{\text{HT}}.
$$
As mentioned above, $\hat{\mu}_{jh}^{\text{HT}}$ represents a Horvitz-Thompson estimate for post-stratum $h$ in small area $j$, and it is weighted by $N_{jh} / N_j$ which is the proportion of population units in small area $j$ that belong to post-stratum $h$. Moreover, the MSE estimator for the post-stratified estimator \parencite{cochran} in small area $j$ is as follows
$$
    \widehat{\text{MSE}}\left(\hat{\mu}_j^{\text{PS}}\right) = \frac{1}{n_j}\left(\sum_{h=1}^2\frac{N_{jh}}{N_j}n_{jh}\widehat{\text{MSE}}\left(\hat{\mu}_{jh}^{\text{HT}}\right) + \sum_{h=1}^2\left(1 - \frac{N_{jh}}{N_j}\right)\frac{n_{jh}}{n_j}\widehat{\text{MSE}}\left(\hat{\mu}_{jh}^{\text{HT}}\right)\right).
$$

\subsubsection{Unit-level EBLUP}\label{sec:unitEBLUP}

Thus far, we have considered direct, design-based estimators. We now turn to the unit-level EBLUP estimator which moves beyond the direct, design-based estimation framework into an indirect, model-based estimation framework, which allows us to leverage data from outside of the small area that we are generating an estimate for, often referred to as ``borrowing strength''. Borrowing strength occurs through the explicit modeling of between-area variation. For the unit-level EBLUP estimator the underlying model is the Battese-Harter-Fuller model, a unit-level, linear mixed model with a random intercept term for small area
\begin{equation} \label{eq:bhf}
    y_{ij} = \mathbf{x}_{ij}^T\boldsymbol{\beta} + \nu_j + \varepsilon_{ij}
\end{equation}
where $\boldsymbol{\beta}$ is the vector of fixed effects regression coefficients, $\nu_j$ is the small area random effect and $\varepsilon_{ij}$ is random noise, distributed as follows
\begin{equation} \label{eq:bhf-error}
     \nu_j \overset{\text{iid}}{\sim} \mathcal{N}(0, \sigma^2_\nu), \quad \varepsilon_{ij} \overset{\text{iid}}{\sim} \mathcal{N}(0, \sigma^2_\varepsilon), \quad \text{and} \quad \nu_j \ind \varepsilon_{ij}.
\end{equation}
Notice that the random intercepts pertain to the small areas in our population and it's these terms that allow us to account for between-area variation. When the model captures the grouped structure of the data through the random effect and through incorporation of valuable auxiliary variables by way of the fixed effects, the unit-level EBLUP estimator can produce more precise estimates than the direct estimators described above. The parameters of this model can be estimated with a restricted maximum likelihood (REML) approach.  

We produce areal estimates by aggregating the unit-level predictions to the small area level. For small area $j$, the unit-level EBLUP estimator of $\mu_j$ is given by
$$
    \hat{\mu}_j^{UEBLUP} = \frac{1}{N_j}\sum_{i\in U_j}\left(\mathbf{X}_{ij}^T\hat{\boldsymbol{\beta}} + \hat\nu_j\right). 
$$
Furthermore, the associated MSE estimator of $\hat{\mu}_j^{EBLUP}$ can be expressed as
$$
    \widehat{\mbox{MSE}}(\hat\mu_j^{UEBLUP}) = g_{1j}(\hat\sigma^2_\nu,~ \hat\sigma^2_\varepsilon) + g_{2j}(\hat\sigma^2_\nu,~ \hat\sigma^2_\varepsilon) + 2g_{3j}(\hat\sigma^2_\nu,~ \hat\sigma^2_\varepsilon)
$$
where $\sigma^2_\nu$ and $\sigma^2_\varepsilon$ are estimated via REML, $g_{1j}$, $g_{2j}$, and $g_{3j}$ capture the within-area variation, variance in estimating regression parameters, and model-variance estimation, respectively. The formal derivations and definitions of $g_{1j}, g_{2j}$, and $g_{3j}$ come from Section 7.2.2 of \textcite{rao15}  as well as \textcite{breidenbach2012small} and are included in Appendix \ref{appendix:equations}

\subsubsection{Area-level EBLUP}

The area-level EBLUP takes a similar linear mixed model approach as the unit-level EBLUP but is fit at the area-level.  It relies on the Fay-Herriot model, which is a linear mixed model with random intercepts on area-level data.  Specifically, the following linking model is assumed:

$$
    \mu_j = \mathbf{\overline{X}}_j^T\boldsymbol{\beta} + \nu_j, \quad \nu_j \overset{\text{iid}}{\sim} \mathcal{N}(0, \sigma^2_\nu).
$$
Further, we assume the following data generation model:
$$
    \mu_j = \hat\mu_j^{\text{HT}} + \varepsilon_j, \quad \varepsilon_j \overset{\text{ind}}{\sim} \mathcal{N}(0, \sigma^2_\varepsilon).
$$
By combining the data generation model and linking model, we obtain the Fay-Herriot model
\begin{equation} \label{eq:fay-herriot}
    \hat\mu_j^{\text{HT}} = \mathbf{\overline{X}}_j^T\boldsymbol{\beta} + \nu_j + \varepsilon_j.
\end{equation}
with the following assumptions 
$$
    \nu_j \overset{\text{iid}}{\sim} \mathcal{N}(0, \sigma^2_\nu) \ , \ \ \  \varepsilon_j \overset{\text{ind}}{\sim} \mathcal{N}(0, \sigma^2_\varepsilon) \ , \ \ \text{and} \ \ \nu_j \ind \varepsilon_j.
$$
We obtain area-level estimates directly from Equation~\ref{eq:fay-herriot}. For small area $j$, the area-level EBLUP estimator of $\mu_j$ is given by
$$
    \hat{\mu}_j^{AEBLUP} = \mathbf{\overline{X}}_j^T\hat{\boldsymbol{\beta}} + \hat{\nu}_j.
$$
Further, our MSE estimator of $\hat{\mu}_j^{AEBLUP}$ is calculated with
$$
    \widehat{\mbox{MSE}}(\hat\mu_j^{AEBLUP}) = f_{1j}(\hat\sigma^2_\nu) + f_{2j}(\hat\sigma^2_\nu) + 2f_{3j}(\hat\sigma^2_\nu)
$$
where $\sigma^2_\nu$ is estimated via REML, and the intuition of the $f_{\#j}$'s are the same as the $g_{\#j}$'s in Section ~\ref{sec:unitEBLUP}. Their derivations and definitions are described in \textcite{rao15} Section 6.2.1. and are also included in Appendix \ref{appendix:equations}.

\subsubsection{Zero-Inflated}

We now formally introduce the zero-inflated small area estimator. In order to specify the estimator we initially specify two models. First, a linear mixed model fit to the nonzero sample data.  Let $s^*_j$ represent the subset of $s_j$ where for $i \in s_j$, it holds that $y_{ij} > 0$ and let 
$s^* = \bigcup_{j=1}^J s^*_j.$  Then we define $y_{ij}^*$ and $\mathbf{x}_{ij}^*$ to be the values of the response and auxiliary variables for the $i$th unit in $s^*_j$, respectively. For the sample data in $s^*$, we fit the following linear mixed model

\begin{equation} \label{eq:zi-lin}
y_{ij}^* = \mathbf{x}_{ij}^{*T}\boldsymbol{\gamma} + u_j + \varepsilon_{ij}
\end{equation}
where $\boldsymbol{\gamma}$ is the vector of fixed effects regression coefficients, $u_j$ is the small area random effect and $\varepsilon_{ij}$ is random noise, distributed as follows:
\begin{equation} \label{eq:zi-lin-error}
     u_j \overset{\text{iid}}{\sim} \mathcal{N}(0, \sigma^2_u), \quad \varepsilon_{ij} \overset{\text{iid}}{\sim} \mathcal{N}(0, \sigma^2_\varepsilon), \quad \text{and} \quad u_j \ind \varepsilon_{ij}.
\end{equation}

In fact, Equations~\ref{eq:zi-lin} and~\ref{eq:zi-lin-error} are analogous to Equations~\ref{eq:bhf} and~\ref{eq:bhf-error} but instead of being fit to the sample data in $s$, they are fit to the sample data in $s^*$. Now, assume that for unit $i$ in small area $j$, $z_{ij}$ follows a Bernoulli distribution where  $p_{ij} = \mathbb{P}(z_{ij} = 1)$.  Using a generalized linear mixed model with a logit link function for $p_{ij}$, we obtain
\begin{equation} \label{eq:zi-log}
    p_{ij} = \frac{1}{1 + \text{exp}\Big(-\big(\mathbf{x}_{ij}^T\boldsymbol{\delta} + w_j\big)\Big)} 
\end{equation}
where $\boldsymbol{\delta}$ is the vector of fixed effects regression coefficients, and the small area random effect $w_j$ is distributed as follows
$$
w_j \sim \mathcal{N}(0, \sigma_w^2).
$$
It is important to note that while Equation~\ref{eq:zi-lin} is fit on only the sample data where the response variable is nonzero, $s^*$, Equation~\ref{eq:zi-log} is fit to the entire sample, $s$. The parameters in Equations~\ref{eq:zi-lin} and~\ref{eq:zi-log} can be estimated using REML.  

Then for $j = 1, \ldots J$ and each $i \in U_j$, we can predict $y_{ij}$ by taking the product of the two fitted models:
$$
    \hat{y}_{ij}^{ZI}  =  \hat{y}_{ij}^* \hat{p}_{ij} = \Big[\mathbf{x}^T_{ij}\hat{\boldsymbol{\gamma}} + \hat{u}_j\Big] \cdot \Bigg[\frac{1}{1 + \text{exp}\big(-\big(\mathbf{x}^T_{ij}\hat{\boldsymbol{\delta}} + \hat{w}_j\big)\big)}\Bigg]
$$
and we can estimate $\mu_j$ by aggregating to the small area level:
$$
    \hat{\mu}_j^{ZI} = \frac{1}{N_j}\sum_{i\in U_j} \hat{y}_{ij}^{ZI}.
$$

We now turn to estimation of the MSE for the zero-inflated estimator. The MSE estimation method that we employ is a parametric bootstrap technique first introduced in \textcite{chandra_sud} which consists of the following steps:

\begin{enumerate}
    \item Generate the bootstrap population.
    \begin{itemize}
        \item Fit the zero-inflated estimator to the original sample data and extract the estimated parameters $\hat{\boldsymbol{\gamma}}, \hat{\boldsymbol{\delta}}, \hat{\sigma}_u^2, \hat{\sigma}_{\varepsilon}^2$, and the realizations of the random effect for the binomial model ($\hat{w}_j$).
        \item  Use $ \hat{\sigma}_u^2$ and $\hat{\sigma}_{\varepsilon}^2$ to generate synthetic area-level random effects and individual random errors by drawing randomly from the following distributions, \begin{equation}\label{eq:syn-area-ranef}
        \tilde{u}_j \sim \mathcal{N}(0, \hat{\sigma}_u^2);\quad (j = 1,~\dots,~J),
        \end{equation}
        and 
        \begin{equation}\label{eq:syn-unit-noise}
        \tilde{\varepsilon}_{ij} \sim \mathcal{N}(0, \hat{\sigma}_{\varepsilon}^2);\quad (i = 1,~\dots,~N_j),~  (j = 1,~\dots,~J).
        \end{equation} 
        Here, we use the tilde ($\sim$) over the quantities generated in Equations~\ref{eq:syn-area-ranef},~\ref{eq:syn-unit-noise}, and throughout the bootstrap process to indicate that these are generated quantities associated with the bootstrap process, not estimated from the sample data. 
        \item Using the estimated parameters obtained from fitting the zero-inflated estimator to the original sample data, calculate the probability of positive value at each population unit, 
        \begin{equation}\label{eq:p-hat-boot}
        \hat{p}_{ij} = \frac{1}{1 + \exp(-(\mathbf{X}_{ij}^T\hat{\boldsymbol{\delta}} + \hat{w}_j))};\quad (i = 1,~\dots,~N_j),~  (j = 1,~\dots,~ J).
        \end{equation}
        \item Generate an indicator variable by drawing from a Bernoulli distribution using the probabilities calculated in Equation~\ref{eq:p-hat-boot}.
        \begin{equation}\label{eq:z-boot}
            \tilde{z}_{ij} \sim \text{Bernoulli}(\hat{p}_{ij});\quad (i = 1,~\dots,~N_j),~  (j = 1,~\dots,~ J).
        \end{equation}
        \item From the quantities generated in Equations~\ref{eq:syn-area-ranef},~\ref{eq:syn-unit-noise}, and~\ref{eq:z-boot}, and $\hat{\boldsymbol{\beta}}$ estimated from the original sample data, generate the response variable for the bootstrap population data,
        \begin{equation}\label{eq:boot-pop}
            \tilde{y}_{ij} = (\mathbf{X}_{ij}^T\hat{\boldsymbol{\beta}} + \tilde{u}_j + \tilde{\varepsilon}_{ij})\cdot \tilde{z}_{ij}\ ;\quad (i = 1,~\dots,~N_j),~  (j = 1,~\dots,~ J).
        \end{equation}
        The full bootstrap population data consists of matching each $\tilde{y}_{ij}$ with the $\mathbf{X}_{ij}$ that was used to generate it.
        $$
        \{\mathbf{X}_{ij}, \tilde{y}_{ij}\} \ ; \quad \left( j = 1, \ldots, J\right), \  \left(i = 1, \ldots N_j\right)
        $$
    \end{itemize}
    \item Calculate synthetic area-level means from the bootstrap population data generated in Equation~\ref{eq:boot-pop},
    \begin{equation}
    \tilde{\mu}_j = \frac{1}{N_j}\sum_{i\in U_j}\tilde{y}_{ij}\ ;\quad(j = 1,~\dots,~ J).
    \end{equation}
    \item Create $B$ bootstrap samples from the bootstrap population data
    \begin{itemize}
        \item To do so, sample $n_j$ units with replacement from each small area in the bootstrap population data $B$ times. It is important emphasize that the sampling is stratified by small area in order to preserve the sample size in each small area.
    \end{itemize}
    \item Fit the zero-inflated estimator on each bootstrap sample to obtain a bootstrap distribution of estimates for the parameter of interest in each small area
    $$
    \hat{\mu}^{(1)}_{j},~\dots,~ \hat{\mu}^{(B)}_{j};\quad(j = 1,~\dots,~ J).
    $$
    \item Obtain MSE estimates using the bootstrap estimates and the synthetic area-level means
    $$
    \widehat{\text{MSE}}(\hat{\mu}_j) = \frac{1}{B}\sum_{b = 1}^B(\hat{\mu}^{(b)}_j - \tilde{\mu}_j)^2;\quad(j = 1,~\dots,~ J).
    $$
\end{enumerate}

\subsection{Simulation Study}\label{sec:sim-study}

We conducted a simulation study set in Nevada to assess the quality of the zero-inflated estimator and its parametric bootstrap MSE estimator. A large majority of the land area of Nevada is classified as desert or other nonforest vegetation cover types. However, isolated mountain ranges known as ``sky islands'' are sparsely distributed throughout the state, and are often substantially forested. Nevada is also a large state at over 110,000 square miles and is divided into just 17 counties. These features make Nevada a great candidate for our simulation study, since (1) we would like to test the zero-inflated estimator in a location where we believe it will be useful, that is, where a significant proportion of the FIA plot data are zero; and (2) the large county size allows us to treat the set of all plots in a given county as the population for that county and then draw a small proportion of those plots for the sample. With this in mind, we now turn to the formal framework of our simulation study.

\subsubsection{Framework}
We let the collection of all FIA plots in Nevada be the population, $U$, of size $N = 11{,}848$. Further, $U$ is partitioned by $J = 17$ counties of size $N_j$, where $j$ indexes over county. We will denote the set containing all FIA plots in a given county by $U_j$. For our simulation study, we create $K = 1000$ Monte Carlo samples from the population $U$, and denote the $k$th sample as $s_k$. In order to generate a given sample, $s_k$, we implement Algorithm~\ref{alg:alg1}. 

\begin{algorithm}
\caption{Creating one sample dataset, $s_k$.}
\begin{algorithmic}
    \State $s_k \gets \emptyset$ 
    \For{$j \in \{1:J\}$} 
        \State{$s_j \gets$ sample($U_j$, proportion = 0.03, replacement = false)}
        \If{$\vert s_j \vert < 2$}
            \State{$s_j \gets$ sample($U_j$, n = 2, replacement = false)}
        \EndIf
        \State{$s_k \gets s_j \cup s_k$}
    \EndFor
\end{algorithmic}
\label{alg:alg1}
\end{algorithm}

Notably, we sample 3\% of FIA plots in each county to give us an average number of FIA plots per county of approximately 20. Further, if sampling 3\% of the county's FIA plots results in a sample of less than two, we sample two plots rather than one or zero. This allows for the within-area variance to be computed for the area-level EBLUP estimator. Repeating Algorithm~\ref{alg:alg1} $K = 1,000$ times gives us the $1,000$ sample datasets we will assess our estimators on.  The minimum number of FIA plots per county across our simulated samples was 2 and the maximum was 57.

\subsubsection{Aims}\label{sec:aims}

Our simulation study aims to assess the zero-inflated estimator in comparison to the EBLUP estimators based on the Battese-Harter-Fuller and Fay-Herriot models, and the post-stratified estimator. In order to make these comparisons, we first introduce our model fitting process, parameters of interest, and metrics for assessing the statistical performance of the estimators.

The estimators are fit in the \texttt{R} language using a variety of packages \parencite{R}. The post-stratified estimator is fit using the \texttt{mase} package \parencite{mase}, the area-level EBLUP is fit using the \texttt{sae} package \parencite{sae}, the unit-level EBLUP is fit using the \texttt{hbsae} package \parencite{hbsae}, and the zero-inflated estimator is fit with our package, \texttt{saeczi}, introduced in Section~\ref{sec:r-package} \parencite{saeczi}. In order to properly and fairly assess the estimators in question, we perform model selection for each estimator across each simulation rep. We use the \texttt{glmnet} package to perform model selection for each estimator on each sample created from the simulation study \parencite{glmnet}. In particular, we use the least absolute shrinkage and selection operator (LASSO) technique for variable selection and select the largest regularization parameter $\lambda$ such that error is within one standard error of the minimum cross validation error. This choice of $\lambda$ puts a greater penalty on the magnitude of the values of the estimated coefficients, leading to a more parsimonious model and a decreased chance of overfitting to the sample data. 

We now turn to discussing the parameters of interest in the simulation study. For a given county $j$, we have two parameters of interest. First 
$$
\mu_j = \frac{1}{N_j} \sum_{i \in U_j} y_{ij},
$$
the average square root biomass in county $j$ and second
\begin{equation} \label{eq:RMSE}
\mbox{RMSE}(\hat\mu_{j\ell}) = \sqrt{\frac{1}{K} \sum_{k=1}^{K} \left(\hat\mu_{jk\ell} - \mu_j\right)^2},
\end{equation}
the empirical root mean square error (RMSE) of the average square root biomass in county $j$, and $\hat\mu_{jk\ell}$ is the estimated mean square root biomass in the $j$th county, $k$th Monte Carlo sample, and $\ell$th estimator. Recall that each estimator we are evaluating produces an estimate of the average square root biomass and an estimate of the RMSE of the estimator for a given county, sample, and estimator. These estimators and their corresponding MSE estimators are defined in Section~\ref{sec:estimators}, and for this analysis we consider the square root of their MSE estimator, the RMSE estimator. 

We now move to discussing the metrics to assess estimator performance. To assess the bias of the estimator for county $j$ and estimator form $l$, denoted $\hat \mu_{jl}$, we compute the percent relative bias (PRB) as follows:
$$
\mbox{PRB}(\hat \mu_{j\ell}) =  \frac{\mathds{E}\left[\hat \mu_{j\ell}\right] - \mu_{j}}{\mu_{j}}  \cdot 100\%
$$
where $\mathds{E}[\hat \mu_{j\ell}]$ is the empirical expected value for estimator $\ell$ of mean square root biomass in county $j$ across all Monte Carlo samples,
\begin{equation} \label{eq:jkl}
    \mathds{E}[\hat \mu_{j\ell}] = \frac{1}{K} \sum_{k=1}^K \hat\mu_{jk\ell}.
\end{equation} 
Beyond bias, we explore the efficiency of the estimators through the county level empirical RMSEs, defined in Equation~\ref{eq:RMSE}.  To determine if our uncertainty estimators are underestimating or overestimating the true uncertainty, we also compute the PRB of the RMSE: 
$$
\mbox{PRB}\left(\widehat{\mbox{RMSE}}\left(\hat \mu_{j\ell}\right)\right) =\frac{\mathds{E}\left[\widehat{\mbox{RMSE}}(\hat\mu_{j\ell})\right] - \mbox{RMSE}(\hat\mu_{j\ell})}{\mbox{RMSE}(\hat\mu_{j\ell})} \cdot 100\%
$$
where $\mathds{E}[\widehat{\mbox{RMSE}}(\hat\mu_{j\ell})]$ is the empirical expected value for estimator $\ell$ of the RMSE for mean square root biomass in county $j$ across all samples defined similarly to Equation~\ref{eq:jkl}. Lastly, we consider the confidence interval coverage probability. We define the 95\% confidence interval coverage for a given county $j$ and estimator $\ell$ as the proportion of samples out of the all $K$ samples where the true parameter, $\mu_j$, falls within the 95\% confidence interval. Formally, we define the estimated confidence interval for the $j$th county, $k$th sample, and $\ell$th estimator form, $C_{jk\ell}$,
$$
C_{jk\ell} = \left[\hat\mu_{jk\ell} - \Phi^{-1}(0.975)\cdot \widehat{\mbox{RMSE}}(\hat\mu_{jk\ell}),~ \hat\mu_{jk\ell} + \Phi^{-1}(0.975)\cdot \widehat{\mbox{RMSE}}(\hat\mu_{jk\ell})\right]
$$
where $\Phi$ is the cumulative density function of the standard normal distribution. Further, we define the indicator function $\mathbbold{1}_{C_{jk\ell}}$,
  $$
    \mathbbold{1}_{C_{jk\ell}}(\mu_j) =
    \begin{cases}
      1, & \text{if}\ \mu_j \in C_{jk\ell}, \\
      0, & \text{if}\ \mu_j \not\in C_{jk\ell}.
    \end{cases}
  $$
The 95\% confidence interval coverage for the $\ell$th estimator form and $j$th county (CICOVG) is thus
$$
\mbox{CICOVG}\left(\hat \mu_{j\ell},~ \widehat{\mbox{RMSE}}\left(\hat \mu_{j\ell}\right)\right) = \frac{1}{K} \sum_{k=1}^{K} \mathbbold{1}_{C_{jk\ell}}.
$$

\section{Results}

\subsection{Simulation Study}\label{sec:sim-res}

We now turn to discussing the results of the simulation study. Here, we examine side-by-side boxplots of each metric described in Section~\ref{sec:aims} in order to assess the performance of the estimators. Because each metric was computed by county, there are 17 values for a given metric and a given estimator.

First, we examine the PRB of the estimator for mean square root biomass in Figure~\ref{fig:PRB}. Here, we see that the zero-inflated estimator's PRB estimates are distributed closely around 0, while the other estimators have more variation in their PRB estimates. Further, the area-level EBLUP estimator is consistently underestimating the average square root biomass in the 17 counties.

The precision of the estimators is summarized in Figure~\ref{fig:RMSE}.  We see that the zero-inflated estimator tends to have a smaller true RMSE for most counties and that unit-level estimation leads to greater precision than area-level estimation. To assess how well the RMSE estimators are quantifying the true uncertainty, we calculate the percent relative bias of the RMSE estimator, shown in Figure~\ref{fig:PRB_RMSE}. All estimators tend to underestimate the true uncertainty and the zero-inflated estimator is competitive with the others. The unit-level EBLUP estimator has much larger variation in its bias, but on median its RMSE estimator is the least biased in our simulation.

Given the underestimation of the uncertainty in our estimates, it is not surprising that there is undercoverage by the 95\% confidence intervals, as shown in Figure~\ref{fig:CI_Coverage}. Here, we again see comparable performance between the unit-level EBLUP estimator and the zero-inflated estimator, with the two other estimators having much lower confidence interval coverage. 

Overall, the zero-inflated estimator tends to have less bias and greater precision than the other estimators and its uncertainty estimator is competitive with the other uncertainty estimators. This indicates that the zero-inflated estimator has much promise and potential to accurately estimate forest inventory metrics such as biomass. In the next section we present the results of the application of these estimators to spatiotemporal domains whose sample sizes are constrained both by geography and by the limited availability of measurements made in a particular year of interest. 

\subsection{Data Application}\label{sec:data-application}

\textcite{rao15} define a ``small area'' as any domain for which direct estimates with adequate precision cannot be produced. In Section \ref{sec:sim-res} the entire land area, forested and non-forested alike, contained by counties in Nevada, considered over a 10-year time period, served as the domains of interest. We subsampled plot measurements made in any year during the associated 10-year FIA inventory cycle using a procedure intentionally devised to yield domain sample sizes small enough to present challenges to both the design- and model-based estimators under investigation. This was particularly convenient as we were able to use the direct estimates for each county across the 10-year FIA inventory cycle as a proxy for truth. 

Now, we turn to applying estimators to an authentic small area problem. In particular, we still consider counties in Nevada as our areas of interest, except now we treat only plots measured in the most recent measurement year (2019) of data as our sample. This approximately reduces our sample size by a factor of 10 and focuses the estimators on estimating average square root biomass in 2019 rather than implicitly assuming domain stasis over the 10-year FIA inventory cycle by incorporating all measurements, irrespective of their measurement year, into domain status estimates. For this data application, only tree canopy cover and elevation were used as auxiliary variables in the models for the model-based estimators, and the binary tree/non-tree variable for the post-stratified estimator. 

Estimates and 95\% confidence intervals for each of the estimators fit are shown in Figure~\ref{fig:DA_estimates}. Notably, the zero-inflated estimator oftentimes has narrower confidence intervals than the other estimators, and sometimes ``splits the difference'' between the area- and unit-level EBLUP estimators (e.g. Storey and Douglas counties). Figure~\ref{fig:DA_chloropleth} displays a chloropleth map of average square root biomass estimates produced by the zero-inflated estimator, which provides a sense of the large-scale spatial variation in biomass across the state of Nevada. Further, Figure~\ref{fig:DA_ref_eff} displays the estimated relative efficiency of each model-based estimator compared to the post-stratified estimator. Notably, two counties have no estimate for relative efficiency due to the post-stratified estimator having an estimated RMSE of zero in those counties. Otherwise, the area-level EBLUP estimator has similar estimated efficiency to the post-stratified estimator, likely due to high variability of direct estimates that are used in the model fitting process for the area-level EBLUP estimator. The unit-level EBLUP estimator and zero-inflated estimator both, on average, show more estimated precision than the post-stratified estimator, however the zero-inflated estimator shows even more estimated precision than the unit-level EBLUP estimator on average. Further, in contrast to the zero-inflated estimator which displays relatively consistent variation across counties, the unit-level EBLUP estimator shows a wide variety of variability and relative efficiency across counties, which can be seen in Figures \ref{fig:DA_estimates} and \ref{fig:DA_ref_eff}. This variation in the unit-level EBLUP estimator's uncertainty estimates is consistent to what we saw in Figure \ref{fig:PRB_RMSE}, where the RMSE estimator for the unit-level EBLUP estimator is nearly unbiased on median, but exhibited a huge variety of bias across counties.  

\section{Discussion}\label{sec:discussion}

We implement the zero-inflated estimator described in \textcite{chandra_sud} within a forest inventory estimation context and compare the efficacy of this estimator to other small area estimators. Further, we introduce the \texttt{R} package \texttt{saeczi} which implements the zero-inflated estimator and its bootstrap MSE estimator through efficient and parallelizable \texttt{R} code and integration with \texttt{C++}. We now turn to formally discussing the \texttt{R} package in Section~\ref{sec:r-package}, and next discuss the implications of our work in Section~\ref{sec:conclusion}. 

\subsection{R Package}\label{sec:r-package}

Here, we introduce \texttt{saeczi}, an \texttt{R} package which implements the zero-inflated estimator and its MSE estimator through efficient programming \parencite{saeczi}. While \textcite{chandra_sud} introduce the zero-inflated estimator and its MSE estimator, they do not provide software to implement the estimator in practice. \texttt{saeczi} implements these estimators through an intuitive function with similar syntax to popular SAE packages in \texttt{R}, informative messaging and progress output as the function runs, and clear and concise examples.  

The implementation of the zero-inflated estimator is a manageable programming task, however efficiently implementing the parametric bootstrap algorithm required for MSE estimation takes more care. Not only is the bootstrap a fairly involved process, but it also is computationally intensive as each iteration requires fitting a zero-inflated estimator on a bootstrap sample and using it to predict on the bootstrap population. Thus, it doesn't take an inordinately large population dataset for the bootstrap to become a very time intensive process under a standard implementation of the algorithm. \texttt{saeczi} allows a user to obtain MSE estimates for a zero-inflated estimator without having to spend a significant amount of time programming the algorithm in an efficient manner. Further, \texttt{saeczi} allows for a user with adequate computational resources to run the estimation process in parallel across threads. 

In addition to the computational efficiency of \texttt{saeczi}'s implementation, we also focused on the user experience. We wrote \texttt{saeczi} to follow design choices of other existing small-area-estimation \texttt{R} packages, such as \texttt{sae} and \texttt{hbsae} \parencite{sae, hbsae}. While the user must still understand the zero-inflated estimator, the alignment of \texttt{saeczi}'s form with existing SAE \texttt{R} packages removes as much friction as possible associated with its use. We hope that using \texttt{saeczi} should feel familiar, and thus integrable, into existing SAE workflows. 

\subsection{Conclusion}\label{sec:conclusion}

Accurate estimates of forest biomass and carbon within small domains are essential for effective forest management, planning, development of sound policy and legislation, and compliance with greenhouse gas emissions reporting requirements. Further, as discussed in Section 5.2 of \textcite{finley2011hierarchical}, much of the world's tree cover is no longer in forests and misclassification of forested and non-forested regions through satellite imagery is substantial. These factors lead us to estimators such as the zero-inflated estimator which are able to accurately quantify average biomass in regions with mixed land cover types (e.g. counties). 

Our work compares the zero-inflated estimator to two common small area estimators, the unit- and area-level EBLUP estimators, and to the post-stratified direct estimator. We achieve these comparisons and assessments through a simulation study and data application, both set in Nevada. The simulation study shows that the zero-inflated estimator has much promise for estimating forest inventory attributes of interest, such as biomass, in areas where levels of zero-inflation are substantial (in this case, we studied small areas where the proportion of zero-valued sample data ranged from 58\% to 98\%). Our work is limited in the sense that we only implement estimators in Nevada, however, we believe the zero-inflated estimator's strong performance will generalize to other similar regions where there are large amounts of land without biomass. In entirely forested regions we do not expect the zero-inflated estimator outperform simpler estimators such as the unit- and area-level EBLUP estimators. 

It's worth noting that the area-level EBLUP estimator is quite different from the other estimation approaches taken in our analyses. We included the area-level EBLUP estimator not only because it is a common indirect model-based approach to small area estimation, but also because it is often used in forest inventory settings where we see model-misspecification in unit-level estimators \parencite{white2021, may2023}. Area-level models exhibit further preferable properties including low computational cost and only requiring access to area-level summaries of inventory data. Area-level models in forest inventory often capture moderately strong linear relationships between the forest attribute of interest and auxiliary data, however this comes at the cost of only being able to leverage a fraction of the available data. Through our simulation study and data application, we show that this trade-off is unfavorable, as the area-level EBLUP estimator consistently produces lower precision (Figure~\ref{fig:RMSE}) and higher bias (Figure~\ref{fig:PRB}) compared to the zero-inflated estimator. In Section~\ref{sec:intro} the zero-inflated estimator was presented as a ``best of both worlds'' approach due to the fact that it takes advantage of the rich unit-level data while simultaneously meeting modeling assumptions. Here we have shown that the additional complexity required to achieve this is indeed worthwhile as the zero-inflated estimator outperforms both the unit- and area-level EBLUP estimators.

One might posit that because the zero-inflated estimator is more complex than the methods used for comparison, i.e., it has more parameters, is a two-stage model, etc., its estimates will improve simply due to the added complexity. However, recent work by \textcite{julianthesis} assessed the same set of estimators alongside a random forest and mixed-effects random forest, in a similar forest inventory setting, and found that the added complexity did not lead to any performance gains over the zero-inflated estimator. Machine learning methods such as random forests can be useful in forest inventory settings, however they often produce unreliable estimates on their own. For instance, \textcite{emick2023} found random forest-produced forest biomass maps to be a useful auxiliary variable in parametric models for forest biomass, but largely unreliable for county-level biomass estimates when used on their own. This is all to say that, while increasing model complexity is a common approach to tackling settings where traditional estimators struggle, it does not always lead to improvements in performance.  The zero-inflated estimator does increase model complexity, but it does so in a way that is entirely motivated by the structure of the data and stays within the class of parametric estimators. It's this approach to adding complexity that allows the zero-inflated estimator to increase precision and accuracy while remaining interpretable. 

In further work, we hope to apply this estimator to even more diverse landscapes and assess it's applicability across the US to understand the sorts of areas this estimator is most useful. Moreover, we hope to assess the uncertainty quantification of the zero-inflated estimators in comparison to other zero-inflated estimators, especially those rooted in the Bayesian paradigm. 

\section*{Author Statements}
\subsection*{Acknowledgements}
The authors thank the Forest Inventory \& Analysis Program for the data, and Romain Boutelet for the French translation of the abstract. 

\subsection*{Competing interests statement}
JKY is employed by RedCastle Resources, Inc. GWW is part-time employed by RedCastle Resources, Inc. The remaining authors declare that the research was conducted in the absence of any commercial or financial relationships that could be construed as a potential conflict of interest.

\subsection*{Funding statement}

This work was supported by the USDA Forest Service, Forest Inventory and Analysis Program, Rocky Mountain Research Station (via agreement 22-JV-11221638-199); the USDA Forest Service, Forest Inventory and Analysis Program, Region 9, Forest Health Protection, Northern Research Station (via agreement 22-CA-11221638-201); by Reed College; by Harvard University; and by Michigan State University AgBioResearch. 

\subsection*{Data availability statement}

The data include confidential plot data, which can not be shared publicly. FIA data can be accessed through the FIA DataMart (\url{https://apps.fs.usda.gov/fia/datamart/datamart.html}). Requests for data used here or other requests including confidential data should be directed to FIA's Spatial Data Services (\url{https://www.fs.usda.gov/research/programs/fia/sds}).

\printbibliography

\section*{Tables}

\begin{table}[H]
 \centering
 \scalebox{0.8}{
 \begin{tabular}{ c|c|c } 
 Variable & Units & Description \\ 
 \hline
 \texttt{ndvi} & index & \makecell{LANDFIRE 2010 Landsat Normalized Difference Vegetation Index \\ \parencite{nelson2015landsat, landfire2016landsat, landfire2020remap}} \\ 
 \texttt{evi} & index  & \makecell{LANDFIRE 2010 Landsat Enhanced Vegetation Index \\ \parencite{nelson2015landsat, landfire2016landsat, landfire2020remap}} \\ 
 \texttt{tcc} & \% & \makecell{National Land Cover Dataset (NLCD) Analytical Tree Canopy Cover \\ \parencite{yang2018new}} \\ 
 \texttt{elev} & meters & \makecell{LANDFIRE 2010 DEM - elevation \\ \parencite{usgs2019ned}} \\  
 \texttt{eastness} & (-100 to 100) & \makecell{Transformed aspect degrees to eastness \\ \parencite{usgs2019ned}} \\ 
 \texttt{northness} & (-100 to 100) & \makecell{Transformed aspect degrees to northness \\ \parencite{usgs2019ned}} \\ 
 \texttt{rough} & meters & \makecell{Degree of irregularity of the surface \\ \parencite{usgs2019ned}} \\ 
 \texttt{tri} & index & \makecell{Terrain Ruggedness Index \\ \parencite{usgs2019ned}} \\ 
 \texttt{tpi} & index & \makecell{Topographic Position Index \\ \parencite{usgs2019ned}} \\ 
 \texttt{ppt} & mm$\times$100 & \makecell{PRISM mean annual precipitation - 30yr normals (1991-2020) \\ \parencite{daly2002knowledge}} \\ 
 \texttt{tmean} & $^{\circ}$C$\times$100 & \makecell{PRISM mean annual temperature - 30yr normals (1991-2020) \\ \parencite{daly2002knowledge}} \\ 
 \texttt{tmax} & $^{\circ}$C$\times$100 & \makecell{PRISM mean annual maximum temperature - 30yr normals (1991-2020) \\ \parencite{daly2002knowledge}} \\ 
 \texttt{tmin} & $^{\circ}$C$\times$100 & \makecell{PRISM mean annual minimum temperature - 30yr normals \\ \parencite{daly2002knowledge}} \\ 
 \texttt{tmin01} & $^{\circ}$C$\times$100 & \makecell{PRISM mean minimum temperature (Jan) - 30yr normals \\ \parencite{daly2002knowledge}} \\ 
 \texttt{tdmean} & $^{\circ}$C$\times$100 & \makecell{PRISM mean annual dewpoint temperature - 30yr normals (1991-2020) \\ \parencite{daly2002knowledge}} \\ 
 \texttt{vpdmax} & Pa$\times$100 & \makecell{PRISM max annual vapor pressure deficit - 30yr normals (1991-2020) \\ \parencite{daly2002knowledge}} \\ 
 \texttt{def} & mm & \makecell{TOPOFIRE mean annual climatic water deficit - 30yr normals (1981-2010) \\ \parencite{holden2019topofire}} \\ 
 \texttt{wc2cl} & class & \makecell{European Space Agency (ESA) 2020 WorldCover global land cover product \\ \parencite{zanaga2021esa}} \\ 
 \texttt{tnt} & class & \makecell{LANDFIRE 2014 tree/non-tree lifeform mask \\ \parencite{rollins2009landfire, picotte2019landfire}}
 \end{tabular}}
 \caption{\label{tab:XVariables}}
\end{table}

\section*{Figure Captions (main body)}
	\begin{singlespace}
Figure 1. Percent relative bias of average square root biomass estimates in each county. The x-axis and the boxplot's fill color indicate the estimator used, while the y-axis indicates the percent relative bias value. This plot shows the percent relative bias across samples for each county. Outliers are labeled by county name and are colored consistently between estimators. 
\\[\baselineskip]
\noindent Figure 2. Percent relative bias of root mean squared error (RMSE) estimators in each county. The x-axis and the boxplot's fill color indicate the estimator used, while the y-axis indicates the percent relative bias value. This plot shows the percent relative bias of the RMSE estimators across samples for each county. Outliers are labeled by county name and are colored consistently between estimators.
\\[\baselineskip]
\noindent Figure 3. Empirical 95\% confidence interval coverage in each county. The x-axis and the boxplot's fill color indicate the estimator used, while the y-axis indicates the empirical 95\% confidence interval coverage. This plot shows the empirical 95\% confidence interval coverage across samples for each county. Outliers are labeled by county name and are colored consistently between estimators.
\\[\baselineskip]
\noindent Figure 4. Empirical root mean squared error (RMSE) values in each county. The x-axis and the boxplot's fill color indicate the estimator used, while the y-axis indicates the empirical RMSE value. This plot shows the empirical RMSE values across samples for each county. Outliers are labeled by county name and are colored consistently between estimators.
\\[\baselineskip]
\noindent Figure 5. Estimates and 95\% confidence intervals of average square root biomass in each county in Nevada for measurement year 2019. Colored dots indicate the estimate and the error bars depict the 95\% confidence interval. Colors distinguish between estimators, with post-stratified estimates in blue, area-level EBLUP estimates in red, unit-level EBLUP estimates in green, and zero-inflated estimates in purple. Counties are shown on the x-axis and are ordered by average estimate value between estimators. 
\\[\baselineskip]
\noindent Figure 6. Estimates of average square root biomass produced by the zero-inflated estimator for each county in Nevada mapped to the polygons of each county. Darker green indicates higher values, while light green indicates lower values. 
\\[\baselineskip]
\noindent Figure 7. Relative efficiency of average square root biomass estimates for each model-based estimator compared to the post-stratified estimator. Yellow values indicate counties that have a lower estimated relative efficiency than the post-stratified estimator and blue values indicate counties with a higher estimated relative efficiency. The left plot shows values for the area-level EBLUP estimator, middle plot for the unit-level EBLUP estimator, and right plot for the zero-inflated estimator. 
	\end{singlespace}

\section*{Figure Captions (Appendix A)}
	\begin{singlespace}
\noindent Figure A1. Quartet of diagnostic plots produced from one sample in the simulation study for the linear mixed model specified for the unit-level EBLUP estimator. The top left plot shows a residual vs. fitted value dot plot with a dotted horizontal line $y = 0$. The top right plot shows a normal QQ plot with a dotted line indicating the $y = x$ line. The bottom left plot shows a scale-location plot with fitted values on the x-axis and the square root of absolute standardized residuals on the y-axis. The blue line is a trend line produced with local polynomial regression fitting. The bottom right plot shows a residual vs. leverage plot with leverage values on the x-axis and standardized residuals on the y-axis. The blue line is a trend line produced with local polynomial regression fitting.
\\[\baselineskip]
\noindent Figure A2. Quartet of diagnostic plots produced from one sample in the simulation study for the linear mixed model specified for the zero-inflated estimator. The top left plot shows a residual vs. fitted value dot plot with a dotted horizontal line $y = 0$. The top right plot shows a normal QQ plot with a dotted line indicating the $y = x$ line. The bottom left plot shows a scale-location plot with fitted values on the x-axis and the square root of absolute standardized residuals on the y-axis. The blue line is a trend line produced with local polynomial regression fitting. The bottom right plot shows a residual vs. leverage plot with leverage values on the x-axis and standardized residuals on the y-axis. The blue line is a trend line produced with local polynomial regression fitting.
	\end{singlespace}

\section*{Figures (main body)}
\begin{figure}[H]
    \includegraphics[width=\textwidth]{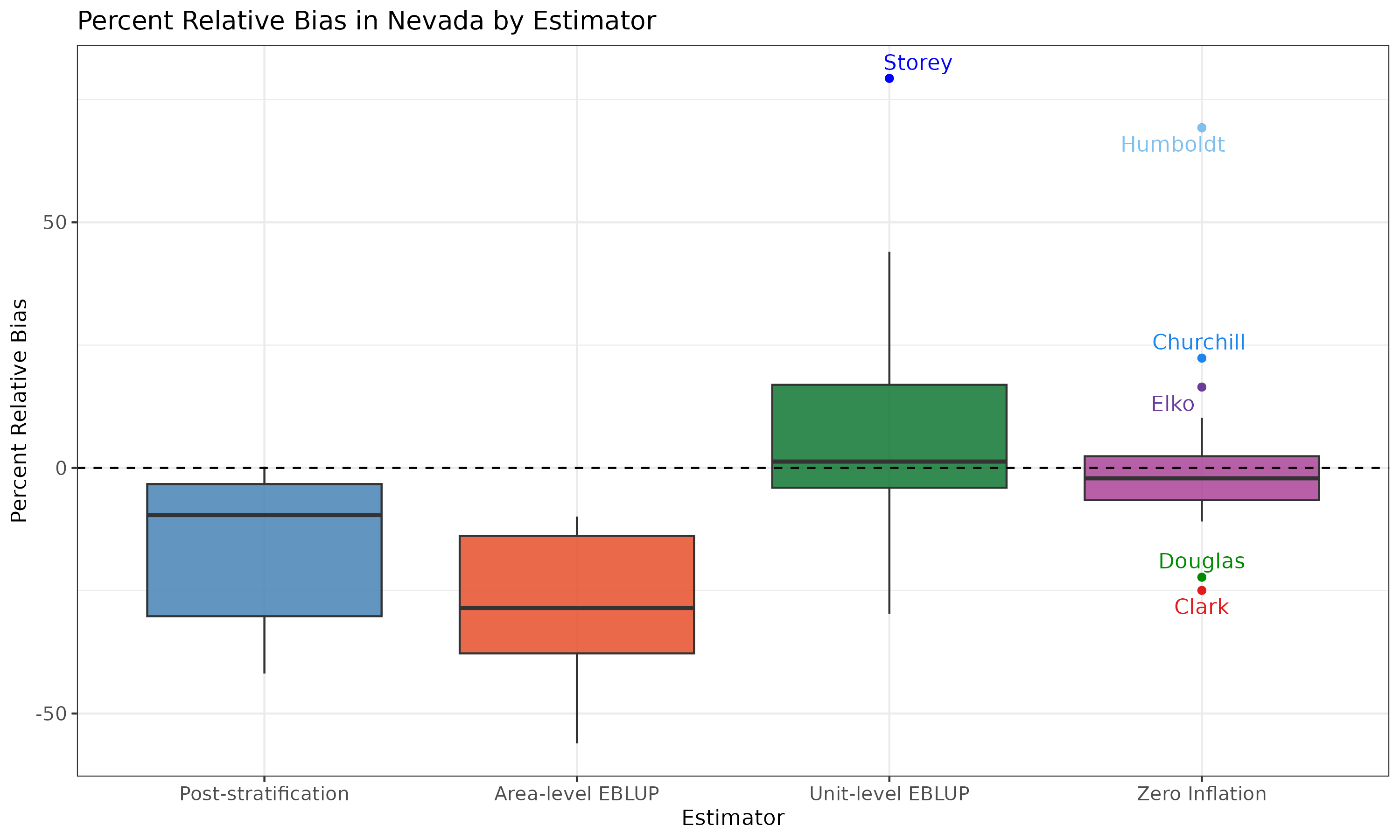}
    \caption{}
    \label{fig:PRB}
\end{figure}

\begin{figure}[H]
    \includegraphics[width=\textwidth]{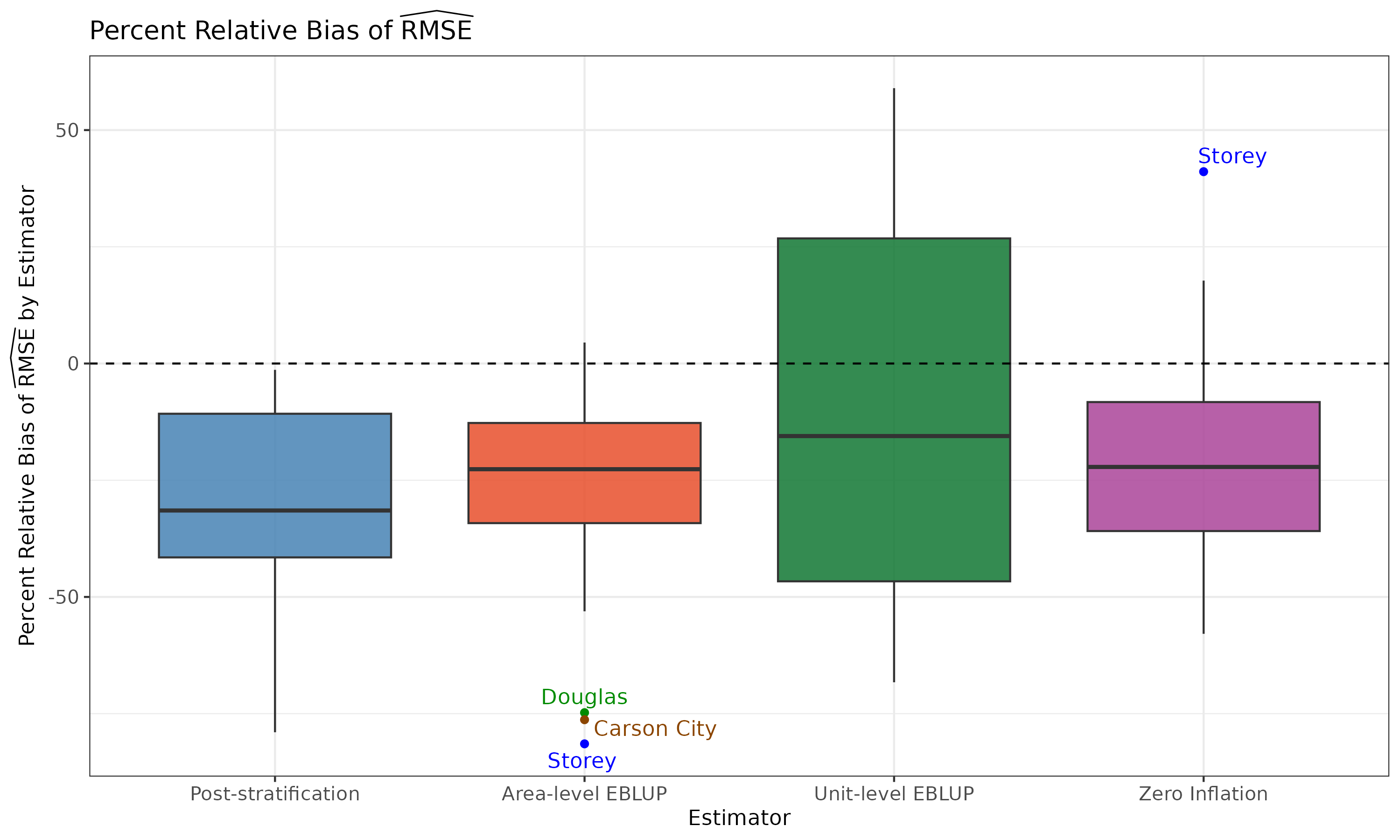}
    \caption{}
    \label{fig:PRB_RMSE}
\end{figure}

\begin{figure}[H]
    \includegraphics[width=\textwidth]{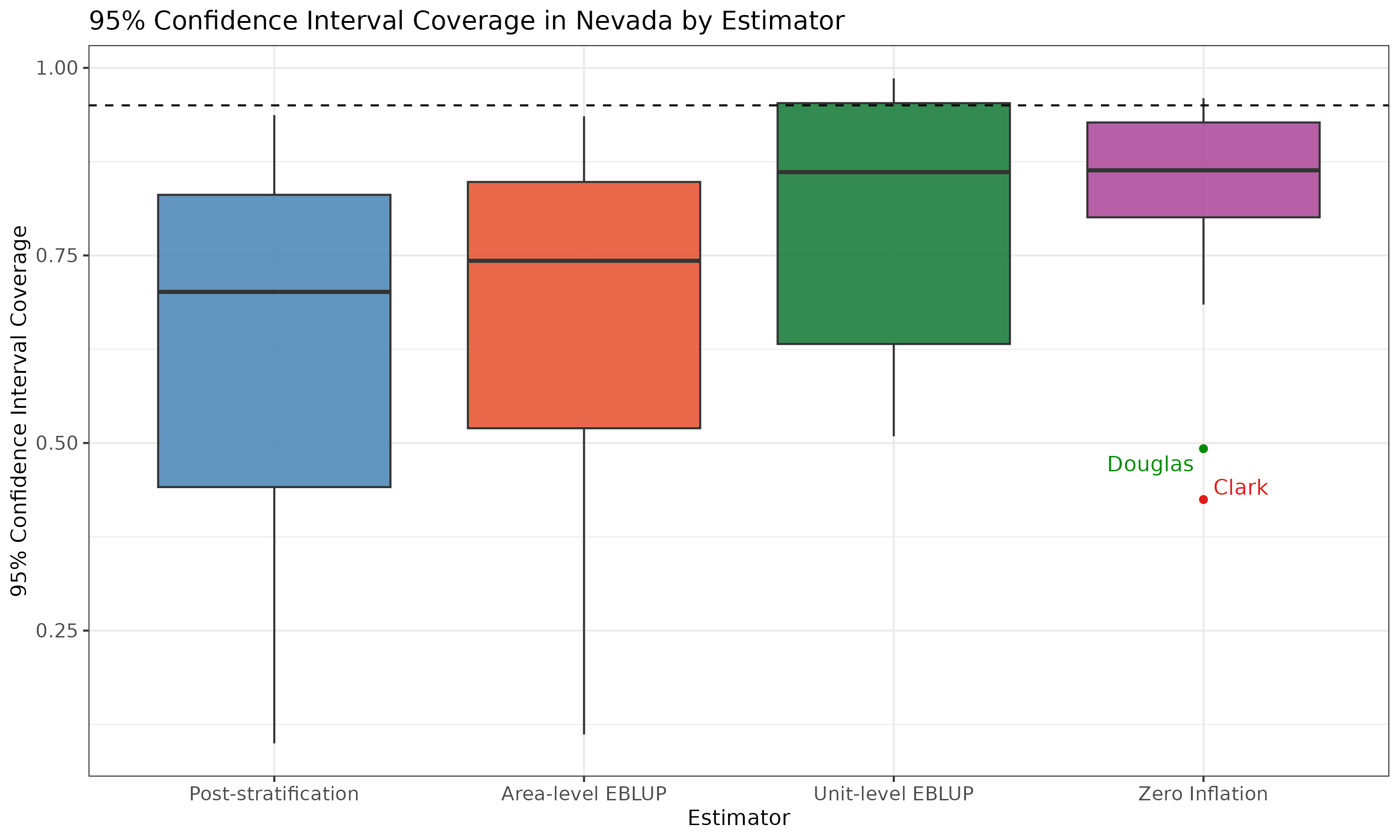}
    \caption{}
    \label{fig:CI_Coverage}
\end{figure}

\begin{figure}[H]
    \includegraphics[width=\textwidth]{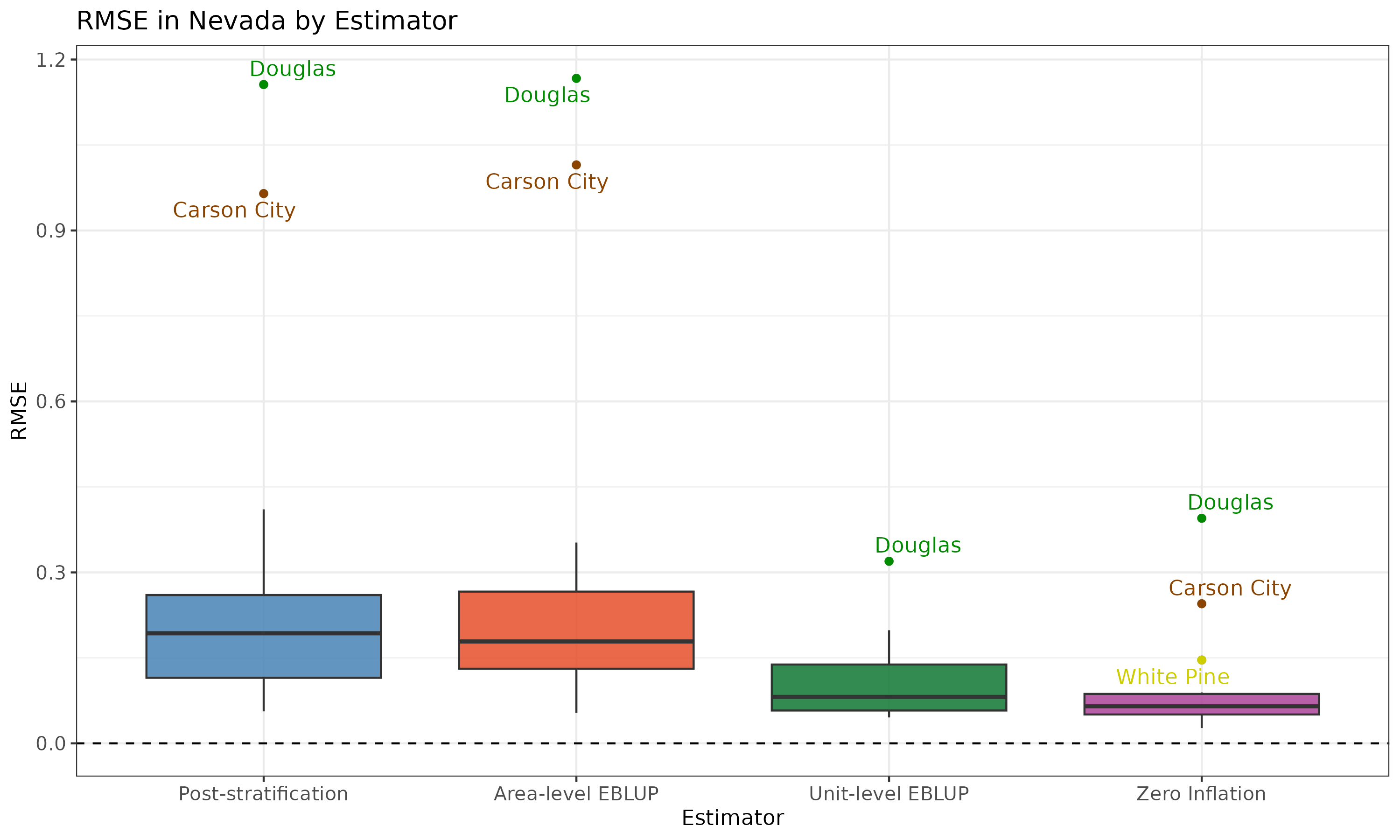}
    \caption{}
    \label{fig:RMSE}
\end{figure}

\begin{figure}[H]
    \includegraphics[width=\textwidth]{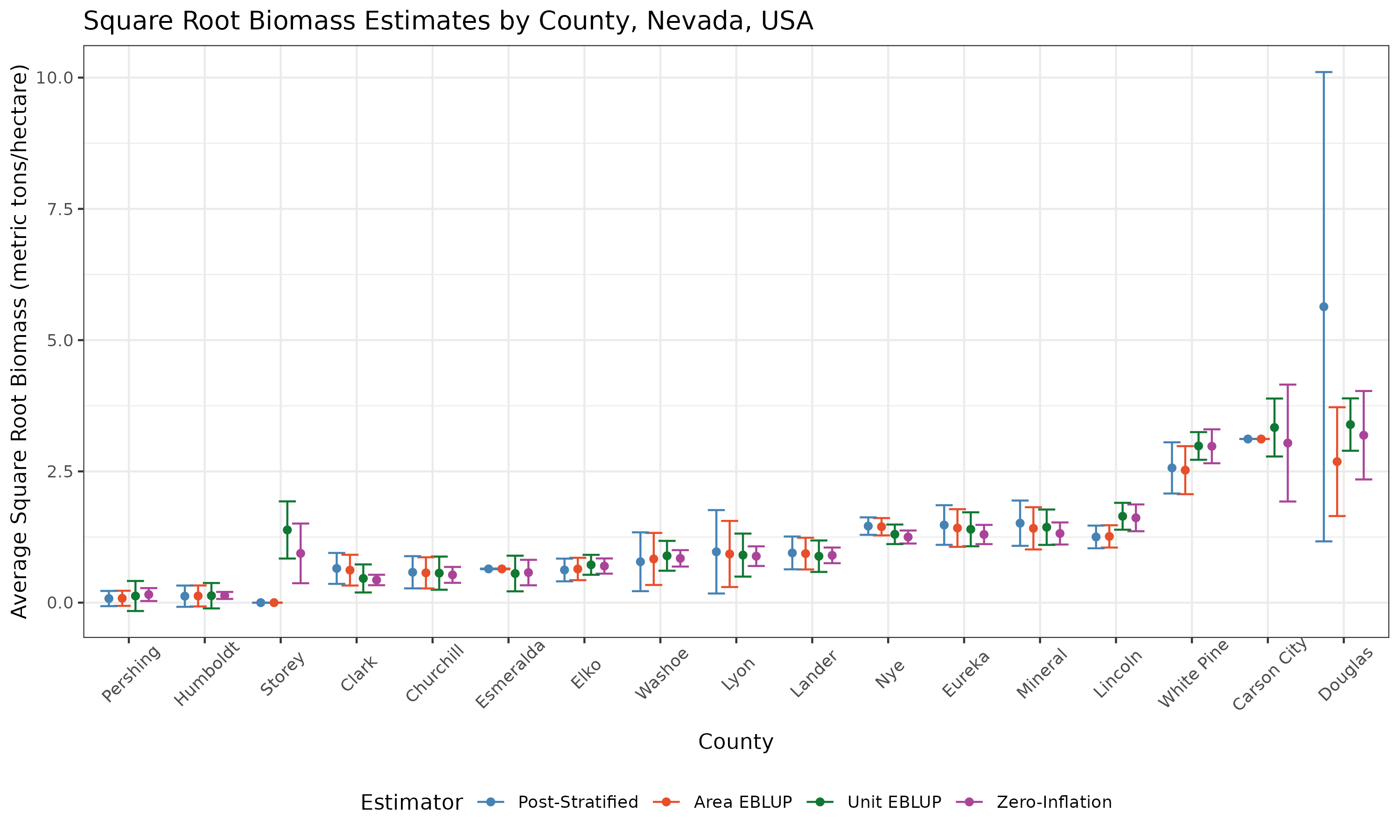}
    \caption{}
    \label{fig:DA_estimates}
\end{figure}

\begin{figure}[H]
    \includegraphics[width=\textwidth]{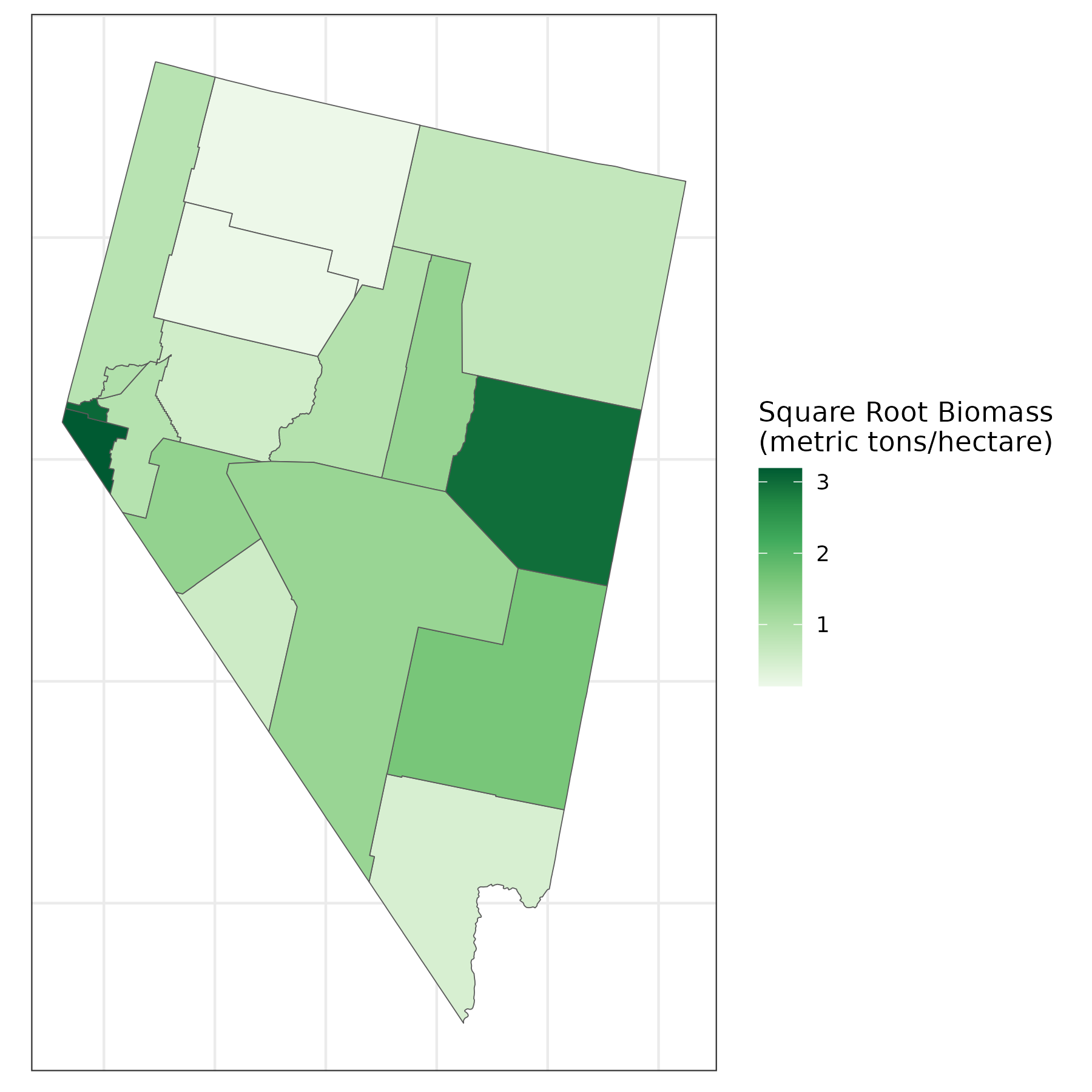}
    \caption{}
    \label{fig:DA_chloropleth}
\end{figure}

\begin{figure}[H]
    \includegraphics[width=\textwidth]{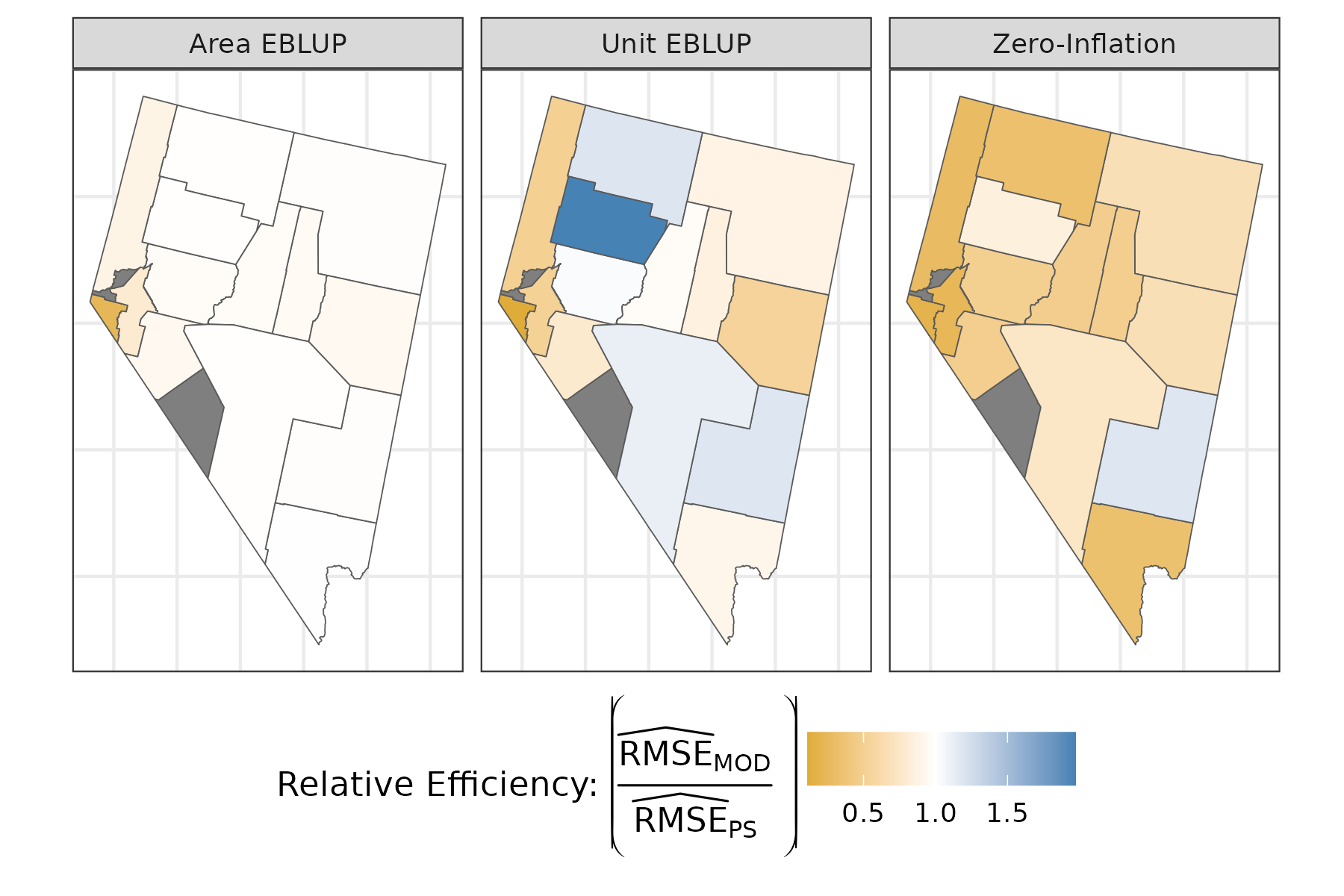}
    \caption{}
    \label{fig:DA_ref_eff}
\end{figure}

\section*{Figures (Appendix A)}
\renewcommand{\thefigure}{A\arabic{figure}}
\setcounter{figure}{0}

\begin{figure}[H]
    \includegraphics[width=\textwidth]{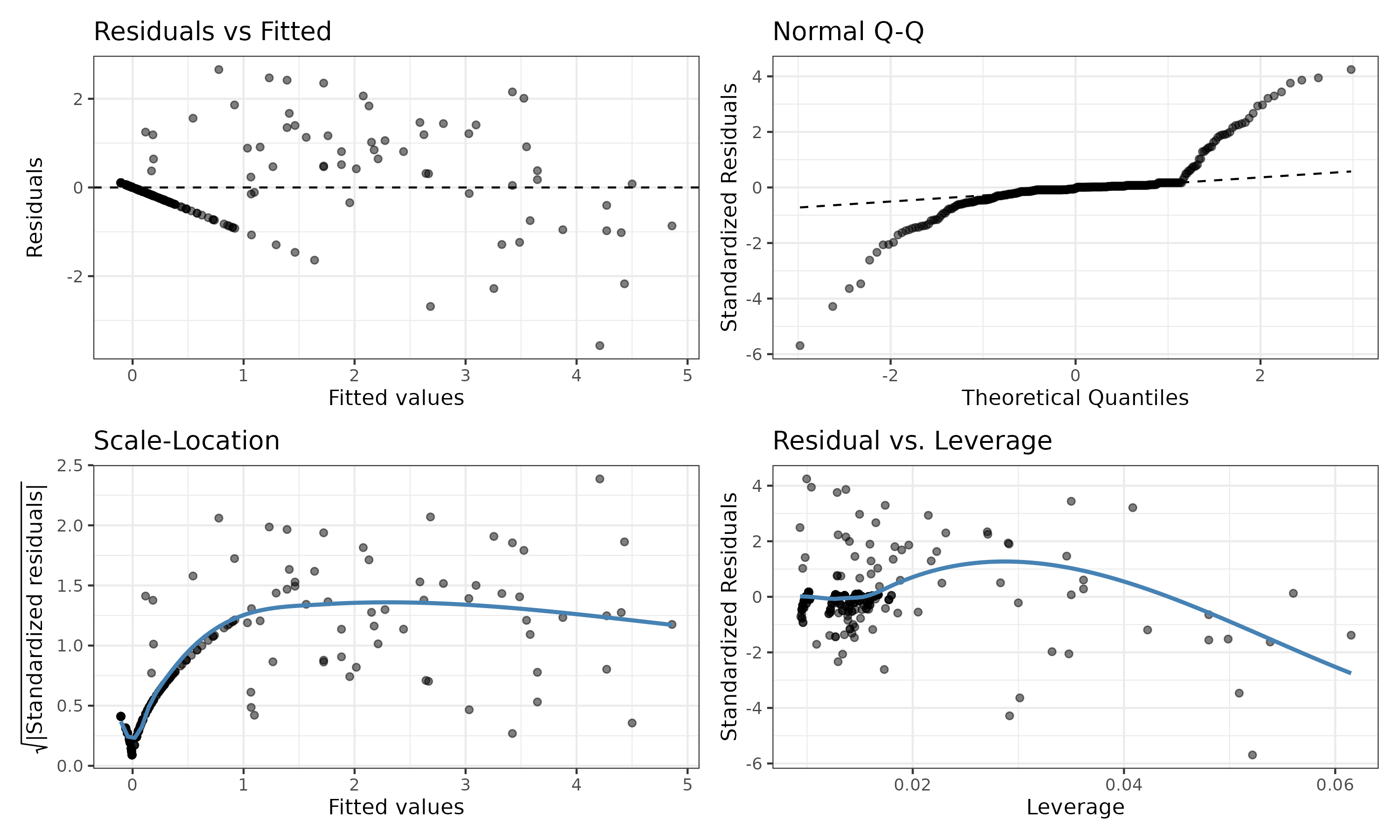}
    \caption{}
    \label{fig:diag_all}
\end{figure}

\begin{figure}[H]
    \includegraphics[width=\textwidth]{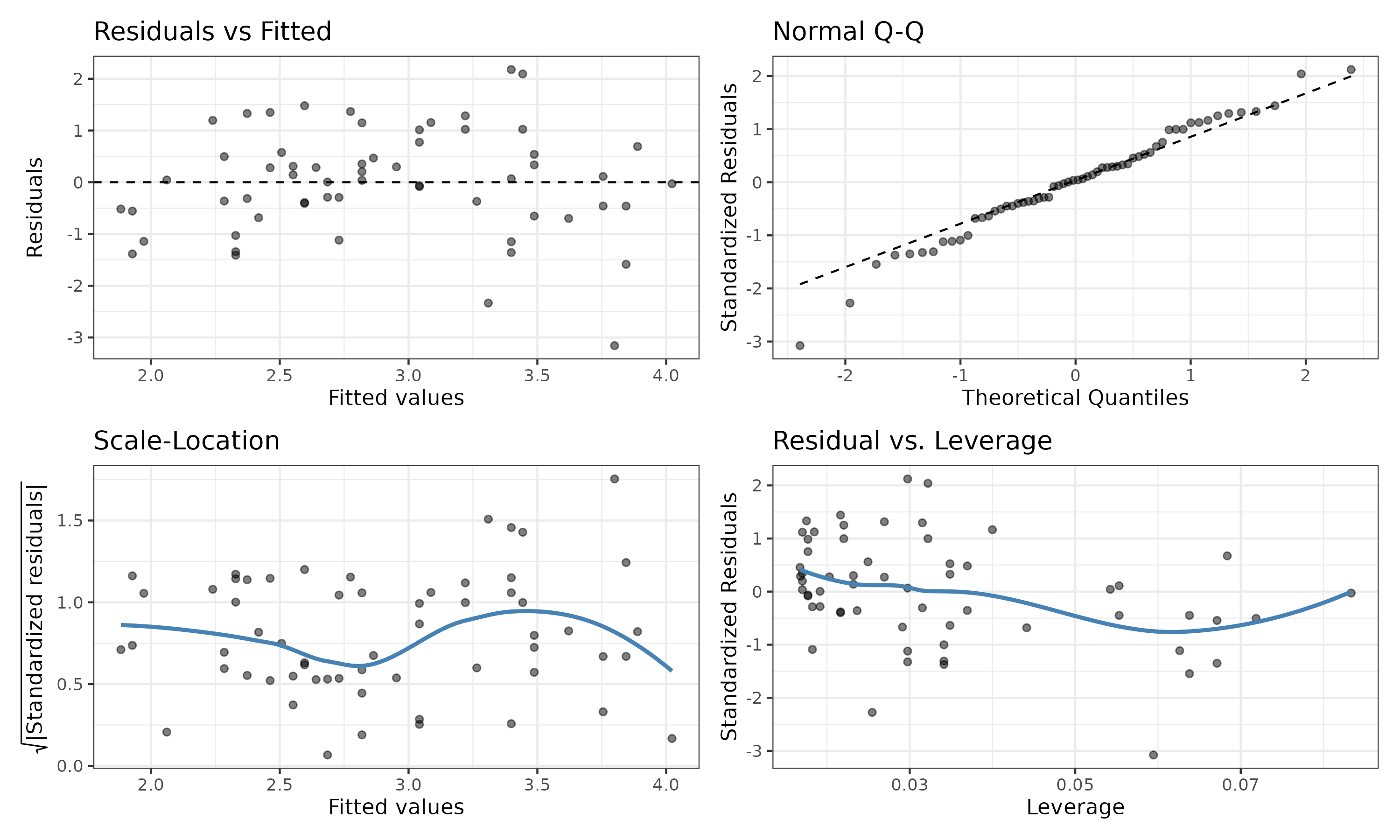}
    \caption{}
    \label{fig:diag_pos}
\end{figure}

\newpage
\appendix
\section*{Appendices}
\section{Diagnostics}
\label{appendix:diagnostics}

For this article, we have focused on estimating average \textit{square root} biomass, due to large amounts of heteroskedasticity when estimating average biomass. In general, we would back-transform to biomass, however the area-level EBLUP estimator produces estimates based on means of pixel-level population data, rather than the pixel-level population data itself, making it impossible to perform the back-transformation. Further, the software used to fit the unit-level EBLUP estimator also uses population means for estimation as a major computational shortcut. Thus, as the goal of this article is primarily to compare estimators under valid modeling assumptions (outside of the zero-inflation in the data, of course), we chose to model the square root biomass. 

When examining the validity of the modeling assumptions through formal diagnostic procedures, the area-level EBLUP estimator generally appeared to meet assumptions. We believe this to be due to the relationship between our response variable and auxiliary data to be much stronger and more linear at the area-level. Further, by aggregating to the area-level, we have entirely removed the issue with zero-inflated plots, as they are absorbed by the average. While choosing an area-level model can be convenient and easy to implement, it also comes with significant losses in performance compared to unit-level alternatives as seen in Figures~\ref{fig:PRB},~\ref{fig:CI_Coverage}, and \ref{fig:RMSE}. Especially in small sample size cases, the area-level EBLUP estimator can perform badly due to unstable direct estimates and direct variance estimates (which are assumed to be a fixed and known quantity in the Fay-Herriot model).

We now turn to discuss modeling assumptions between the unit-level model-based estimators. Recall that the unit-level EBLUP estimator fits a linear mixed model to the sample data, but the zero-inflated estimators fits a linear mixed model on the \textit{positive} sample data. This distinction is a large motivation for the zero-inflated estimator, and by examining common diagnostic plots we can see why the model fit on the positive sample data is much more sensible than the one fit on the entire sample data. For one sample from the simulation study using just tree canopy cover and elevation as linear predictors, Figure~\ref{fig:diag_all} displays the quartet of modeling diagnostic plots for the linear mixed model fit on the entire sample data (i.e. the Battese-Harter-Fuller model), and Figure~\ref{fig:diag_pos} displays the quartet of modeling diagnostic plots for the linear mixed model fit on the positive sample data (i.e. the linear model fit by the zero-inflated estimator).

The modeling assumptions are clearly violated in Figure~\ref{fig:diag_all}. Most notably, the residual vs fitted plot trends downwards with a heavy diagonal line caused by the zero-valued plots, the normal-QQ plot has extremely heavy tails, and the scale-location plot suffers from a distinctive ``check mark'' of points caused by the zero-values plots. However, these issues are largely corrected in Figure~\ref{fig:diag_pos}. Here, we see a much more sensible residuals vs. fitted plot where the points uncorrelated with each other and normally distributed around zero, the points on the normal-QQ plot generally fall on the theoretical line, and there are no patterns in the scale-location plot. This significant improvement of modeling diagnostics strongly motivates the zero-inflated estimator, and showcases the pitfalls of using a unit-level EBLUP estimator when the sample data contains zero-valued cases, but is otherwise positive and continuous. 

\section{EBLUP Estimated MSE Equations}
\label{appendix:equations}

\subsection*{Area-level}

The following equations represent the terms for the estimated MSE for the area-level EBLUP estimator \parencite{rao15}.

\begin{align*}
    f_{1j}(\hat{\sigma}_{\nu}^2) &= \hat{\gamma}_j\widehat{\mbox{MSE}}\left(\hat{\mu}_j^{HT}\right) \\
    f_{2j}(\hat{\sigma}_{\nu}^2) &= \left(1 - \hat{\gamma}_j\right)^2 \mathbf{\overline{X}}_j^T\left[\sum_{j=1}^J\mathbf{\overline{X}}_j\mathbf{\overline{X}}_J^T / \left(\widehat{\mbox{MSE}}\left(\hat{\mu}_j^{HT}\right) + \hat{\sigma}_{\nu}^2\right)\right]^{-1}\mathbf{\overline{X}}_j \\
    f_{3j}(\hat{\sigma}_{\nu}^2) &= \widehat{\mbox{MSE}}\left(\hat{\mu}_j^{HT}\right)^2\left(\widehat{\mbox{MSE}}\left(\hat{\mu}_j^{HT}\right) + \hat{\sigma}_{\nu}^2\right)^{-3} \left[2\sum_{j=1}^J 1 / \left(\hat{\sigma}_{\nu}^2 + \widehat{\mbox{MSE}}\left(\hat{\mu}_j^{HT}\right)\right)^2\right]^{-1}
\end{align*}

where 

$$
    \hat{\gamma}_j = \hat{\sigma}_{\nu}^2 / \left(\widehat{\mbox{MSE}}\left(\hat{\mu}^{HT}\right) + \hat{\sigma}_{\nu}^2\right)
$$

\subsection*{Unit-level}

The following equations represent the terms for the estimated MSE for the unit-level EBLUP estimator \parencite{rao15, breidenbach2012small}.

\begin{align*}
    g_{1j}(\hat{\sigma}_{\nu}^2, \hat{\sigma}_{\varepsilon}^2) &= 
    \hat{\gamma}_j \left(\hat{\sigma}_{\nu}^2 / n_j\right) \\
    g_{2j}(\hat{\sigma}_{\nu}^2, \hat{\sigma}_{\varepsilon}^2) &= 
    \left(\overline{\mathbf{X}}_j -\hat{\gamma}_j\overline{\mathbf{x}}_j\right)^T\left[\sum_{j=1}^{J}\hat{\sigma}_{\varepsilon}^{-2}\left(\sum_{i = 1}^{n_j}\mathbf{x}_{ij}y_{ij} - \hat{\gamma}_j n_j\overline{\mathbf{x}}_j\overline{y}_j\right)\right]^{-1} \left(\overline{\mathbf{X}}_j - \hat{\gamma}_j\overline{\mathbf{x}}_j\right) \\
    g_{3j}(\hat{\sigma}_{\nu}^2, \hat{\sigma}_{\varepsilon}^2) &= n_j^{-2}\left(\hat{\sigma}_{\nu}^2  + \hat{\sigma}_{\varepsilon}^2 / n_j\right)^{-3}h\left(\hat{\sigma}_{\nu}^2, \hat{\sigma}_{\varepsilon}^2\right)
\end{align*}

where 

\begin{align*}
    \hat{\gamma}_j &= \hat{\sigma}_{\nu}^2 / \left(\hat{\sigma}_{\nu}^2 + \hat{\sigma}_{\varepsilon}^2 / n_j \right)
\end{align*}

and

\begin{align*}
    h\left(\hat{\sigma}_{\nu}^2, \hat{\sigma}_{\varepsilon}^2\right) &= \hat{\sigma}_{\varepsilon}^4\left[1/2 \sum_{j=1}^Jn_j^2 \left(\hat{\sigma}_{\varepsilon}^2 + n_j\hat{\sigma}_{\nu}^2\right)^{-2}\right]^{-1} + \hat{\sigma}_{\nu}^4\left[1/2 \sum_{j=1}^J\left(\left(n_j-1\right)\hat{\sigma}_{\varepsilon}^{-4} + \left(\hat{\sigma}_{\varepsilon}^2 + n_j\hat{\sigma}_{\nu}^2\right)^{-2}\right)\right]^{-1} \\ 
    & \quad - 2\hat{\sigma}_{\varepsilon}^2\hat{\sigma}_{\nu}^2\left[1/2 \sum_{j = 1}^Jn_j\left(\hat{\sigma}_{\varepsilon}^2 + n_j \hat{\sigma}_{\nu}^2\right)^{-2}\right]
\end{align*}

\end{document}